\documentclass[aps,prb,twocolumn,amsmath,amssymb,footinbib,showpacs,longbibliography,superscriptaddress]{revtex4-1}

\usepackage[english]{babel}
\usepackage{latexsym}
\usepackage{graphics}
\usepackage{subfigure}
\usepackage{epsfig}
\usepackage{color}
\usepackage{hyperref}
\usepackage{braket} 
\usepackage[T1]{fontenc}
\usepackage[latin9]{inputenc}
\usepackage{amstext}
\usepackage{amssymb}
\usepackage{graphicx}
\usepackage{esint}
\usepackage{braket}
\usepackage{babel}
\usepackage{amsmath}
\hypersetup{
colorlinks=true,
citecolor=blue,
linkcolor=red,
urlcolor=black
}


\global\long\def\R{\mathbf{R}}
\global\long\def\Rp{\mathbf{R}^{\prime}}
\global\long\def\k{\mathbf{k}}

\newcommand{\Vg}{V_{\textrm{g}}}
\newcommand{\Vphi}{V_{\varphi}}
\newcommand{\VLR}{V_{\textrm{\tiny CE}}}
\newcommand{\Vloc}{V_{\textrm{m}}}

\newcommand{\kmax}{k^{\star}}
\newcommand{\ksp}{k_{\textrm{sp}}}
\newcommand{\Vgmax}{\Vg^{\star}}
\newcommand{\Vphimax}{\Vphi^{\star}}

\newcommand{\betaLR}{{\beta_{\textrm{\tiny CE}}}}
\newcommand{\betam}{{\beta_{\textrm{\tiny m}}}}
\newcommand{\betaLRfit}{{\beta_{\textrm{\tiny CE}}^{\textrm{fit}}}}
\newcommand{\betamfit}{{\beta_{\textrm{\tiny m}}^{\textrm{fit}}}}

\newcommand{\e}{\textrm{e}}
\newcommand{\ie}{\textit{i.e.}}
\newcommand{\eg}{{e.g.}}

\newcommand{\Jnature}{Nature (London)}
\newcommand{\Jnatphys}{Nat. Phys.}
\newcommand{\Jnatcomm}{Nat. Comm.}
\newcommand{\Jnatmat}{Nat. Mater.}

\newcommand{\Jscience}{Science}

\newcommand{\Jprl}{Phys. Rev. Lett.}
\newcommand{\Jpr}{Phys. Rev.}
\newcommand{\Jpra}{Phys. Rev. A}
\newcommand{\Jprb}{Phys. Rev. B}

\newcommand{\Jprx}{Phys. Rev. X}

\newcommand{\Jnjp}{New J. Phys.}

\newcommand{\Jjchemphys}{J. Chem. Phys.}

\newcommand{\JRepProgPhys}{Rep. Prog. Phys.}

\newcommand{\JjphysB}{J. Phys. B: At. Mol. Opt. Phys.}

\newcommand{\JCommMathPhys}{Comm. Math. Phys.}

\begin{document}

\title{
Universal scaling laws for correlation spreading in quantum systems with short- and long-range interactions
}

\author{Lorenzo Cevolani}
\affiliation{
 Institut f\"ur Theoretische Physik,
 Georg-August-Universit\"at G\"ottingen, 37077 G\"ottingen, Germany
}

\author{Julien Despres}
\affiliation{
 Centre de Physique Th\'eorique, Ecole Polytechnique, CNRS, Univ Paris-Saclay, F-91128 Palaiseau, France
}

\author{Giuseppe Carleo}
\affiliation{
 Institute for Theoretical Physics,
 ETH Zurich,
 Wolfgang-Pauli-Str.~27,
 8093 Zurich, Switzerland
}

\affiliation{
 Center for Computational Quantum Physics, Flatiron Institute, 162 5th Avenue, New York, NY 10010, USA
}

\author{Luca Tagliacozzo}
\affiliation{
 Department of Physics and SUPA,
 University of Strathclyde,
 Glasgow G4 0NG,
 United Kingdom
}

\author{Laurent Sanchez-Palencia}
\affiliation{
 Centre de Physique Th\'eorique, Ecole Polytechnique, CNRS, Univ Paris-Saclay, F-91128 Palaiseau, France
}

\date{\today}

\begin{abstract}
The spreading of correlations after a quantum quench is studied in a wide class of lattice systems, with short- and long-range interactions.
Using a unifying quasi-particle framework, we unveil a rich structure of the correlation cone,
which encodes the footprints of several microscopic properties of the system.
When the quasi-particle excitations propagate with a bounded group velocity, we show that the correlation edge and correlation maxima move with different velocities that we derive.
For systems with a divergent group velocity, especially relevant for long-range interacting systems, the correlation edge propagates slower than ballistic.
In contrast, the correlation maxima propagate faster than ballistic in gapless systems but ballistic in gapped systems.
Our results shed new light on existing experimental and numerical observations, and pave the way to the next generation of experiments.
For instance, we argue that the dynamics of correlation maxima can be used as a witness of the elementary excitations of the system.
\end{abstract}

\maketitle

\section{Introduction} The ability of a quantum system to establish long-distance correlations and entanglement, and possibly equilibrate, is determined by the speed at which information can propagate within the system. For lattice models with short-range interactions, Lieb and Robinson (LR) have unveiled a bound that forms a linear causality cone beyond which information decays exponentially~\cite{lieb1972}.
This bound implies ballistic propagation of equal time-correlation functions~\cite{bravyi2006} that has been observed experimentally~\cite{cheneau2012,fukuhara2013} and characterized numerically~\cite{manmana2009,barmettler2012,carleo2014,dellanna2016,kormos2017}. 
Generalized LR bounds have been derived for long-range systems where the interactions decay algebraically, $1/R^\alpha$, with the distance $R$, see Ref.~\onlinecite{hastings2006,foss-feig2015}. The related experiments and  numerical investigations have, however, lead to conflicting pictures~\cite{hauke2013,eisert2013,jurcevic2014,richerme2014,cevolani2015,cevolani2016,buyskikh2016}.
For instance, experiments on ion chains~\cite{richerme2014} and numerical simulations within truncated Wigner approximation~\cite{schachenmayer2015b}
for the one-dimensional (1D) long-range XY (LRXY) model
point towards bounded, super-ballistic, propagation for all values of $\alpha$. In contrast, experiments on the long-range transverse Ising (LRTI) model reported ballistic propagation of correlation maxima with, however, observable leaks that increase when $\alpha$ decreases~\cite{jurcevic2014}.
Moreover, time-dependent density matrix renormalization group (t-DMRG) and variational Monte-Carlo (t-VMC) numerical simulations indicate
the existence of three distinct regimes, namely instantaneous, sub-ballistic, and ballistic, for increasing values of the exponent $\alpha$, see Ref.~\onlinecite{schachenmayer2013,hauke2013,eisert2013,cevolani2015,cevolani2016,buyskikh2016}.

In this paper we shed light on these apparent contradictions.
We focus on equal-time correlation functions that are relevant experimentally.
Implications of bounded correlation spreading on universal LR bounds are not yet completely understood (see however Ref.~\onlinecite{kastner2015})
so we do not draw explicit conclusions of the latter.

Using a universal picture based on quasi-particles that can be applied to both short- and long-range models, we unveil a double causality structure for correlation spreading. The \emph{outer structure} determines the correlation edge (CE), while \emph{the inner structure} determines the propagation of local extrema.
For short-range interactions, the two structures are  determined by the
dispersion relation and can be associated to, respectively, the group and phase velocities of the quasi-particles.
For long-range interactions, the inner structure is still determined by the dispersion relation. It is super-ballistic for gapless models and ballistic for gapped models. It implies that quantum quenches can be used experimentally as a witness
to detect the presence of the gap and the value of the dynamical exponent of the underlying model, something that as far as we know was not realized previously.  
The outer structure depends both on the dispersion relation and on the considered observable, and is thus less universal. Besides pathologic cases, it is always sub-ballistic. 

The identification of this double structure (\ie\ edge versus local maxima of correlations), and the lack of universality of the outer-edge in long-range systems permits to accomodate and explain previous observations in a unified picture. 
In particular it constitutes an important result to predict the spreading of specific observables and design the next generation of experiments within a large class of  long-range systems,
\eg\
Rydberg gases~\cite{bendkowsky2009,weimer2010,schausz2012,browaeys2016},
nonlinear optical media~\cite{firstenberg2013},
polar molecules~\cite{micheli2006,yan2013,moses2017}, magnetic atoms~\cite{griesmaier2005,beaufils2008,lu2011,baier2016,lahaye2009},
superconductors~\cite{houck2012},
ion chains~\cite{deng2005,islam2011,lanyon2011,schneider2012b,NaturePhysicsInsight2012blatt},
and solid-state defects~\cite{childress2006,balasubramanian2009,dolde2013}.

\section{Time evolution of local correlations}
Consider a quantum system defined on a hypercubic lattice of dimension $D$ and governed by a translation-invariant Hamiltonian of the form
%%%%%%%%%%%%%%%%%%%%%%%%%%%%%%%%%%
\begin{equation}
\label{eq:H}
\hat{H} = \sum_{\R} h (\R)\, \hat{K}_1({\R}) + \sum_{\R,\R'} J (\R,\R')\, \hat{K}_2(\R,\R'),
\end{equation}
%%%%%%%%%%%%%%%%%%%%%%%%%%%%%%%%%%
where $\R$ and $\Rp$ span the lattice sites. The first term accounts for local interactions and the second term for two-site couplings.
It applies to a variety of models including the Bose-Hubbard (BH), see App.~\ref{sec:appBH}, the LRXY, and the LRTI models, see App.~\ref{sec:LRModels}, that we consider in the following.
We start from the ground-state of $\hat{H}$ and quench the system out of equilibrium by changing the couplings at time $t=0$.
We  characterize the evolution by computing equal-time connected correlation functions with respect to the pre-quench equilibrium value. They read $G (\R,t) \equiv G_0\left( \R, t\right)-G_0\left( \R, 0\right)$ with $G_0\left( \R,t \right)\equiv \langle \hat{A}_{X}(t) \hat{B}_{Y}(t)\rangle-\langle \hat{A}_{X}(t)\rangle\langle \hat{B}_{Y}(t)\rangle$,
where $\hat{A}_X$ and $\hat{B}_Y$ are local operators with support on regions $X$ and $Y$ separated by $\R$. Such correlations can be measured in state-of-the-art experiments~\cite{cheneau2012,fukuhara2013,richerme2014,jurcevic2014}.
We describe quenches where the dynamics is driven by the low-energy sector of $\hat{H}$ that may be assumed to consist of quasi-particle excitations. Due to  translation invariance, they are characterized by well-defined quasi-momentum $\k$ and energy $E_{\k}$. The correlation functions may be written
%%%%%%%%%%%%%%%%%%%%%%%%%%%%%%%%%%
\begin{equation}
\label{eq:gen-corr-tev}
G\!\left(\R,t\right) \! = \! g(\R) -
\int_{\mathcal{B}}\frac{d\k}{\left(\!2\pi\!\right)^{\!D}}\mathcal{F}\left(\k\right)\frac{\e^{i\left(\k\cdot\R+2E_{\k}t\right)} \! + \! \e^{i\left(\k\cdot\R-2E_{\k}t\right)}}{2},
\end{equation}
%%%%%%%%%%%%%%%%%%%%%%%%%%%%%%%%%%
where the integral spans the first Brillouin zone $\mathcal{B}$.
The quantity $g(\R)$ can be dropped, since it does not depend on time.
Equation~(\ref{eq:gen-corr-tev}) represents the motion of counter-propagating quasi-particle pairs, with velocities determined by  $E_{\k}$, and where the amplitude $\mathcal{F}(\k)$ encodes the overlap of the initial state with the  quasi-particle wave functions and the matrix elements of $\hat{A}$ and $\hat{B}$.
It can be derived explicitly in exactly-solvable models and quadratic systems, which can be diagonalised by means of canonical transformations.
Many models, in various regimes, can be mapped into this form
(see for instance Refs.~\onlinecite{barmettler2012,natu2013,hauke2013,cevolani2015,cevolani2016,buyskikh2016,frerot2018} in the context of out-of-equilibrium dynamics). The concept of quasi-particles also applies to non exactly-solvable models, where they can be determined using tensor-network techniques~\cite{haegeman2012,nakatani2014} for 
instance, and we expect that our results also hold for such systems.

\section{Short-range couplings}
Consider first the case of nearest-neighbour interactions
for which the quasi-particle group velocity is bounded.
In the infinite time and distance limit along the line $R/t=\textrm{const}$, the integral in Eq.~(\ref{eq:gen-corr-tev}) is dominated by the momentum contributions with a stationary phase (sp), \ie\ $\nabla_{k} (k R \mp 2E_k t )=0$ or, equivalently,
%%%%%%%%%%%%%%%%%%%%%%%%%%%%%%%%%%
\begin{equation}
\label{eq:statphase1}
2\Vg (\ksp) = \pm R/t,
\end{equation}
%%%%%%%%%%%%%%%%%%%%%%%%%%%%%%%%%%
where $\Vg=\nabla_k E_k$ is the group velocity.
Since the latter is upper bounded by some value $\Vgmax$,
Eq.~(\ref{eq:statphase1}) has a solution only for $R/t<2 \Vgmax$. The correlation function then reads~\cite{SeveralMax}
%%%%%%%%%%%%%%%%%%%%%%%%%%%%%%%%%%
\begin{equation}
\label{eq:statphase2}
G(R,t) \propto \frac{\mathcal{F}(\ksp)}{\big(\vert\nabla_{k}^{2}E_{\ksp}\vert t \big)^{\frac{D}{2}}}\cos\left(\ksp R-2E_{\ksp}t + \frac{\pi}{4} \right).
\end{equation}
%%%%%%%%%%%%%%%%%%%%%%%%%%%%%%%%%%
For $R/t>2 \Vgmax$, Eq.~(\ref{eq:statphase1}) has no solution and $G(R,t)$ is vanishingly small.
The correlations are thus activated ballistically at the time $t=R/2\Vgmax$. It defines a linear correlation edge (CE) with velocity $\VLR=2\Vgmax$,
consistently with the Calabrese-Cardy picture~\cite{calabrese2006}.
% Note that for $\kmax \neq 0$, the quantity $\mathcal{F}(\kmax)$ is in general finite while $\vert \nabla_{k}^{2}E_{\kmax}\vert$ vanishes.
% Hence the amplitude of $G(R,t)$ diverges on the cone, which indicates a sharp CE.

Yet, Eq.~(\ref{eq:statphase2}) does not only yield the CE but also a series of local maxima. In the vicinity of the CE cone, only the quasi-particles with momenta $k \simeq \kmax$, which move at $\Vgmax$, contribute to the correlations. There the maxima (m), defined by the equation $\kmax R-2E_{\kmax}t=\textrm{const}$, propagate at the velocity $\Vloc=2\Vphimax \equiv 2E_{\kmax} / \kmax$, \ie\ twice the phase velocity at the maximum of the group velocity, $\kmax$.
Since the phase and group velocities are generally different, 
the CE is expected to feature a double structure characterized by these two velocities.
This observation and its counterpart for long-range systems (see below) have fundamental consequences on correlation spreading and is the pivotal result of this work.

\begin{figure}[t!]
\includegraphics[width = \columnwidth]{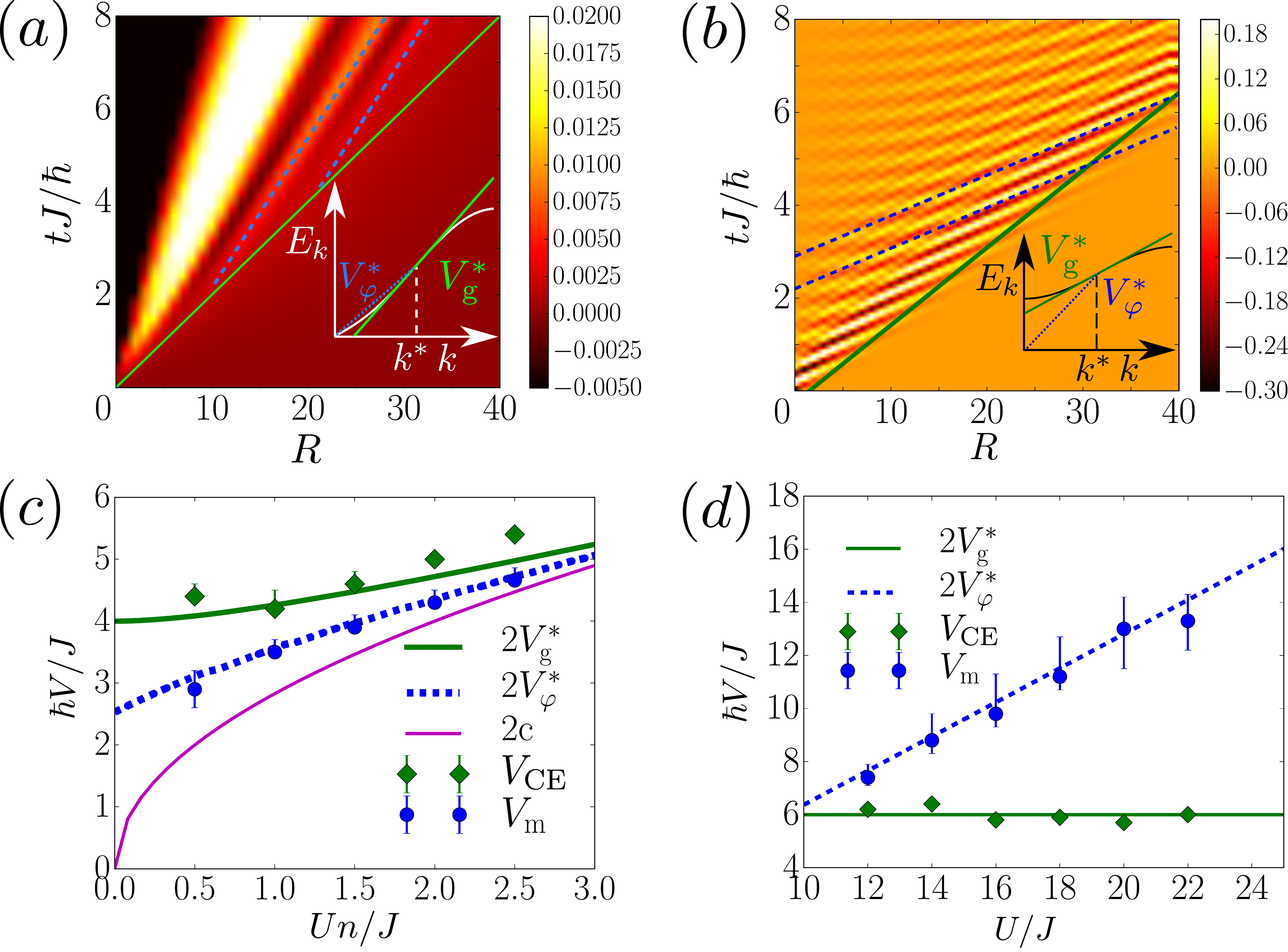}
\caption{\label{fig:BHm}
Upper panel:~Spreading of the connected one-body correlation function
$G(R,t)=\langle \hat{a}^\dagger_R (t) \hat{a}_0 (t) \rangle - \langle \hat{a}^\dagger_R (0) \hat{a}_0 (0) \rangle$
for the 1D Bose-Hubbard model.
(a)~Superfluid phase for a quench from the initial value $U_\textrm{i}n=J$ to the final value $U_\textrm{f}n=0.5J$.
(b)~Mott-insulator phase with $n=1$ for a quench from $U_\textrm{i}=\infty$ to $U_\textrm{f}=18J$.
The solid green and dashed blue lines indicate ballistic spreading at twice the maximum group velocity, $2\Vgmax$, and twice the corresponding phase velocity, $2\Vphimax$, respectively.
Lower panel:~Comparison between the maximum group velocity ($\Vgmax$, solid green line), the corresponding phase velocity ($\Vphimax$, dashed blue line), the sound velocity ($c$, dotted purple line),
and fits to the LR cone velocity ($\VLR$, green diamonds) and to the velocity of the maxima ($\Vloc$, blue disks) for the (c)~superfluid and (d)~Mott insulator phases with the same initial values as for (a) and (b).
}
\end{figure}

To illustrate it, let us consider the BH model.
In the superfluid regime, the dispersion relation is bounded and the group velocity has a local maximum at some momentum $0<\kmax<\pi$, see inset of Fig.~\ref{fig:BHm}(a).
%%%%%%%%%%%%%%%%%%%%%%%%%%%%%%%%%%
%%%%%%%%%%%%%%%%%%%%%%%%%%%%%%%%%%
The main panel of Fig. 1(a) shows the connected one-body correlation function versus distance and time in this regime. Its value is determined from numerical integration of Eq.~\eqref{eq:gen-corr-tev} with the coefficients calculated using Bogoliubov theory.
The latter holds for weak interactions, $Jn \gg U$, with $J$ the hopping and $U$ the interaction strength, see App.~\ref{sec:appBH} for details.
As expected, the correlation cone is determined by the velocity $\VLR \simeq 2\Vgmax$ (solid green line). Moreover, the correlations show a series of local maxima, all propagating at the same speed, approximately twice the phase velocity at the momentum $\kmax$, $\Vloc \simeq 2\Vphimax$ (dashed blue lines).
These observations are confirmed quantitatively in Fig.~\ref{fig:BHm}(c) where we compare the values of the velocities found from fits to the correlation edge ($\VLR$) and local maxima ($\Vloc$) on the one hand, to twice the group ($\Vgmax$) and phase ($\Vphimax$) velocities at $\kmax$ on the other hand.

The distinction between the edge propagating at $\Vgmax$ and the maxima emanating from it and propagating at $\Vphimax$ permits to understand previously unexplained observations. The propagation velocity extracted from t-VMC calculations in Ref.~\onlinecite{carleo2014} quantitatively agrees with the value $2\Vphimax$ calculated here. Our analysis shows that it should thus be assimilated to the propagation of the local maxima, \ie\  \emph{the inner structure} of the causal region~\cite{NoDoubleInPrev}.
In contrast, the CE is determined by the raise of the envelop of these maxima, and moves at the velocity $2\Vgmax$. The same analysis applies to the results of t-DMRG calculations for a quench in the superfluid regime of the Fermi-Hubbard model~\cite{manmana2009}.

So far experimental characterization of correlation spreading for a quench in the Mott-insulator regime was  performed close to the critical point where there is a single characteristic velocity and as a consequence no inner structure was observed~\cite{cheneau2012}.
A richer behaviour occurs deeper in the Mott regime, where the presence of a gap
permits to find a regime, $U>\pi(2n+1)J$,
where $\Vphimax>\Vgmax$, see inset of Fig.~\ref{fig:BHm}(b).
Here the excitation spectrum is found using strong-coupling perturbation theory, which holds for $n\in\mathbb{N}^*$ and $U \gg Jn$, see App.~\ref{sec:appBH}. The connected one-body correlation function plotted in Fig.~\ref{fig:BHm}(b) is found from numerical integration of Eq.~(\ref{eq:gen-corr-tev}) with the corresponding coefficients.
In this case, the local maxima propagate (still at $\Vphimax$) faster than the correlation cone (still at $\Vgmax$), and vanish when reaching it, see Figs.~\ref{fig:BHm}(b) and (d).

\section{Long-range couplings}
We now turn to long-range systems with power-law couplings, $J_{\R,\Rp} \sim J/\vert\R-\Rp\vert^\alpha$. We assume that the spectrum is regular in the whole Brillouin zone, except for a cusp at $k=0$.
There, the dispersion relation may be written $E_{k} \simeq \Delta+ck^z$, with $z$ the dynamical exponent and $\Delta$ the (possibly vanishing) gap. For $0<z<1$ the quasi-particle energy $E_k$ is bounded but the group velocity $\Vg(k)$ diverges.
In the following, we consider connected spin correlation functions for two spin models, as found from Eq.~(\ref{eq:gen-corr-tev}) and linear spin wave theory, see App.~\ref{sec:LRModels} and references therein. All quenches are performed in a single polarized phase, without crossing any critical line.

\begin{figure}
  \centering
  \includegraphics[width = \columnwidth]{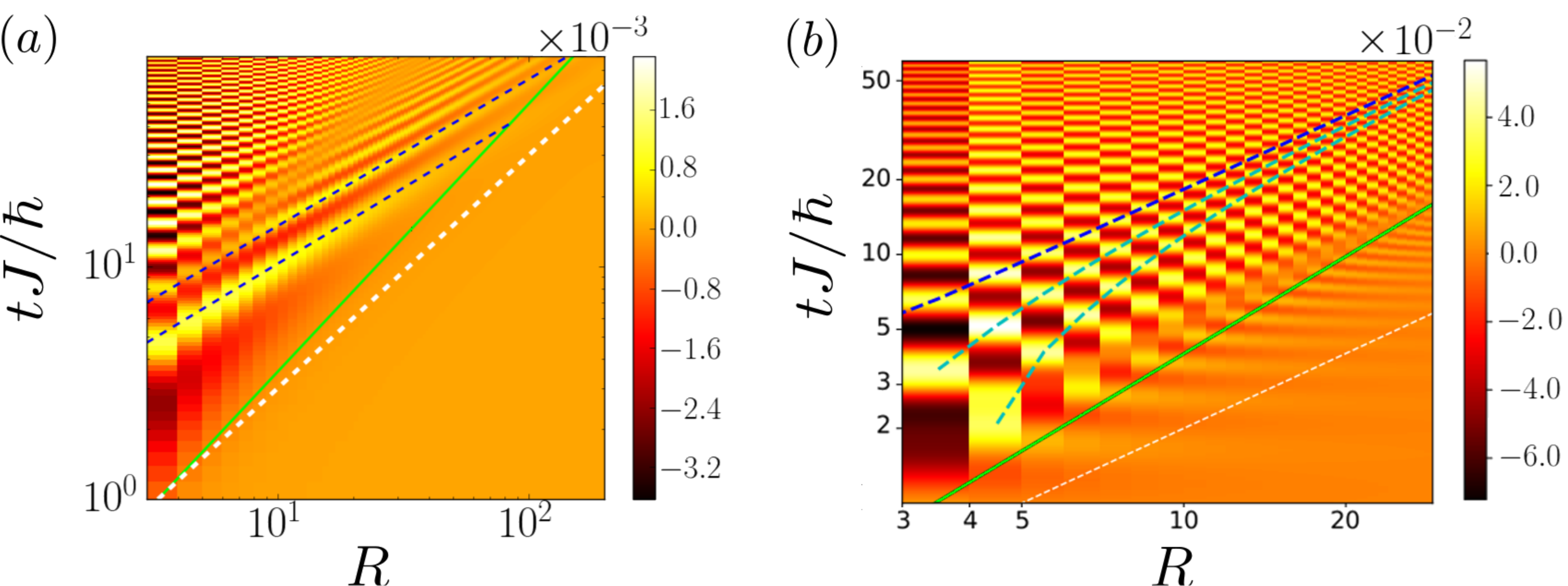}
  \caption{
  Spreading of the connected spin correlation function $G(R,t)=G_0(R,t)-G_0(R,0)$
for the following 1D models:
  (a)~LRXY model with $\alpha=2.3$, $G_0(R,t) = \langle S_R^z(t) S_0^z(t) \rangle - \langle S_R^z(t)\rangle \langle S_0^z(t) \rangle$ for a quench from the ground state of the XXZ model ($\epsilon=0.2$, see App.~\ref{sec:LRModels});
  (b)~LRTI model with $\alpha=1.7$, $G_0(R,t) = \langle S_R^x(t) S_0^x(t) \rangle - \langle S_R^x(t)\rangle \langle S_0^x(t) \rangle$ for the quench in the polarized phase, from $J_{\textrm{i}}/h=0.02$ to $J_{\textrm{f}}/h=1$, see Ref.~\onlinecite{LRTItransition}.
  They feature a double algebraic structure (straight lines in log-log scale):
  A sub-ballistic correlation edge (solid green line) and super-ballistic or ballistic spreading of local maxima (dashed blue lines).
  The white dotted line indicates ballistic spreading for reference. The light blue, dashed lines are guides to the eye.
}\label{fig:spins}
\end{figure}

Figure~\ref{fig:spins}(a) corresponds to the LRXY model.
Owing to continuous spin rotation symmetry,
it is gapless, $\Delta=0$, and $z=(\alpha-D)/2$ for $D<\alpha<D+2$, see Ref.~\onlinecite{frerot2017}.
Figure~\ref{fig:spins}(b) corresponds to the LRTI model, where the transverse magnetic field opens a gap, $\Delta>0$, and $z=\alpha-D$ for $D<\alpha<D+1$, see Ref.~\onlinecite{hauke2013,cevolani2015,cevolani2016}.
For both models, we find a double structure reminiscent of the one of short-range models, although with crucial differences. 
First, the CE is not linear but algebraic (note the log-log scales in Fig.~\ref{fig:spins}). 
While the known extended LR bounds~\cite{hastings2006,foss-feig2015} are all super-ballistic, we find a \textit{sub-ballistic} CE, $t \sim R^\betaLR$ with $\betaLR>1$
(the edges are marked by solid green lines and, for reference, ballistic spreading by white dotted lines).
Second, the inner structure shows a strongly model-dependent behaviour:
For the LRXY model [Fig.~\ref{fig:spins}(a)], the correlation maxima (dashed blue lines) are \textit{super-ballistic}, $t \sim R^\betam$ with $\betam<1$ while for the LRTI model [Fig.~\ref{fig:spins}(b)] they are \textit{ballistic}, $t \sim R$.

%%%%%%%%%%%%%%%%%%%%%%%%%%%%%%%%%%
%%%%%%%%%%%%%%%%%%%%%%%%%%%%%%%%%%

To understand these behaviours, let us use again the stationary-phase approximation. Equations~(\ref{eq:statphase1}) and (\ref{eq:statphase2}) still hold.
However, the group velocity, $\Vg(k)=|c|z/k^{1-z}$, now diverges at $k \to 0$. Hence, for any combination of $t$ and $R$, there is a quasi-particle with the corresponding group velocity, at the momentum $\ksp=\left(2|c|zt/ R\right)^{1/\left(1-z\right)}$.
The CE is thus dominated by the infrared divergence, where we now need to analyse the amplitude function $\mathcal{F}$. Inserting  the assumed scaling $\mathcal{F}(k) \sim k^\nu$, with $\nu \geq 0$,  into Eqs.~(\ref{eq:statphase1}) and (\ref{eq:statphase2}), we find
%%%%%%%%%%%%%%%%%%%%%%%%%%%%%%%%%%
\begin{equation}
\label{eq:statphaseLR}
G_{\textrm{c}}(R,t) \propto \frac{t^{\gamma}}{R^{\chi}}\cos\left[A_z\left(\frac{t}{R^{z}}\right)^{\frac{1}{1-z}}-2\Delta t + \frac{\pi}{4}\right],
\end{equation}
%%%%%%%%%%%%%%%%%%%%%%%%%%%%%%%%%%
with
    $\gamma=\frac{\nu+D/2}{1-z}$,
    $\chi=\frac{\nu+D(2-z)/2}{1-z}$, and
   $A_{z}=2|c| (1-z) (2|c|z)^{\frac{z}{1-z}}$.
The CE is found by imposing that the amplitude of the correlation function becomes of order one. It yields to the algebraic form
%%%%%%%%%%%%%%%%%%%%%%%%%%%%%%%%%%
\begin{equation}
\label{eq:LRfront}
t^{\star} \propto R^{\betaLR},
\qquad \betaLR=\chi/\gamma.
\end{equation}
%%%%%%%%%%%%%%%%%%%%%%%%%%%%%%%%%%
Hence, the scaling of the CE does not depend only on the dynamical exponent $z$ but also on the specific correlation function, via the exponent $\nu$, and on the dimension $D$. This contrasts with the short-range case, where a ballistic propagation independent of the dimension and of the observable is found~\cite{ShortRangeLimit}.
Since $\chi=\gamma+D/2$ the CE is always sub-ballistic, $\betaLR>1$.
For the LRXY model and spin-spin correlations, we have
$\nu=z=(\alpha-D)/2$,
which yields $\betaLR=
1+\frac{D}{2\alpha}\left(2+D-\alpha\right)$.
In the numerical calculations of Fig.~\ref{fig:spins}(a), the CE is found by tracing the points in the $R-t$ plane where the correlations reach $\epsilon=2\%$ of the maximal value.
The activation time $t^{*}$ as a function of the distance $R$ is then fitted by a power law, $t^{*} \sim R^{\betaLRfit}$, see details in App.~\ref{sec:Edgeextraction}.
For $\alpha=2.3$ and $D=1$ [Fig.~\ref{fig:spins}(a)], we find $\betaLRfit \simeq 1.083 \pm 0.013$, in good agreement with the theoretical value $\betaLR \simeq 1.15$.

For the LRTI model, we have $z=\alpha-D$ and $\nu=0$.
It yields the exponent $\betaLR=2-z=2+D-\alpha$ completely determined by the dynamical exponent $z$.
For $\alpha=1.7$ and $D=1$ [Fig.~\ref{fig:spins}(b)], analyzing the numerical results as before
we find the CE exponent $\betaLRfit = 1.28 \pm 0.02$,
in excellent agreement with the theoretical value $\betaLR = 1.3$.
Note that the general formula for $\betaLR$ matches the exact result of Ref.~\onlinecite{cevolani2015} for $D=1$ and $\alpha=3/2$ (\ie\ $z=1/2$), also confirmed by t-VMC calculations, and it is in fair agreement with the analysis of Ref.~\onlinecite{cevolani2016} for the 1D and 2D LRTI models.

On the other hand, the inner structure of the causal region is determined by the local maxima of the cosine function in Eq.~(\ref{eq:statphaseLR}).
It does not depend on the observable but on the presence or absence of a gap.
For a gapless system ($\Delta=0$), we find
%%%%%%%%%%%%%%%%%%%%%%%%%%%%%%%%%%
\begin{equation}
\label{eq:LRnogapmax}
t_\textrm{m} \propto R^{\betam},
\qquad \betam=z.
\end{equation}
%%%%%%%%%%%%%%%%%%%%%%%%%%%%%%%%%%
The correlation maxima are thus always super-ballistic, $\betam<1$.
For the LRXY model and $\alpha=2.3$ [Fig.~\ref{fig:spins}(a)], we find the theoretical value $\betam=0.65$.
In the numerics, we study the internal structure of the correlation function by tracking the position of the first local maximum as a function of time. We then fit
the corresponding function by $t_\textrm{m} = a R^{\betamfit}+b$. For the parameters of Fig.~\ref{fig:spins}(a), it yields $\betamfit \simeq 0.634 \pm 0.014$, in excellent agreement with the theoretical value.
It is also consistent with the experimental observation of super-ballistic dynamics in the 1D LRXY model realized with trapped ion chains for $\alpha>1$, see Ref.~\onlinecite{richerme2014} and in
rough agreement with the analysis of numerical calculations performed within the truncated Wigner approximation for 1D and 2D LRXY models~\cite{schachenmayer2015b}.
The same result as Eq.~(\ref{eq:LRnogapmax}) was found in Ref.~\onlinecite{frerot2018}, which appeared while completing the present work.
Our analysis shows that this super-ballistic behaviour characterizes the inner structure but not the CE.

For a gapped system ($\Delta >0$), the momentum dependence of the dispersion relation becomes irrelevant in the infrared limit and the argument of the cosine function in Eq.~(\ref{eq:statphaseLR}) is constant in the large $t$ and $R$ limit for $t \propto R$.
It follows that the local maxima are here always ballistic, $\betam=1$.
This case applies to the LRTI model.
It is confirmed in Fig.~\ref{fig:spins}(b), where we observe that the local maxima converge to a ballistic propagation for sufficiently long times.
Performing the analysis as above, we find $\betamfit \simeq 1.0045 \pm 0.0003$, in excellent agreement with the theoretical prediction.
This result is consistent with the observation of ballistic motion of local maxima for the 1D LRTI model realized with trapped ion chains~\cite{jurcevic2014}.

\section{Conclusions}
In this work we have shown that the spreading of equal-time correlations has a double structure
whose scaling laws can be related to different characteristic spectral properties.
For short-range systems, they are readily associated to the group and phase velocities, which generally differ.
For long-range systems with a diverging group velocity,
the CE is observable-dependent and sub-ballistic.
Close to the CE, the local maxima propagate ballistically in gapped systems and super-ballistically in gapless systems.
Their observation can thus be used as an experimental footprint for the presence of a spectral gap.

This double structure can be observed experimentally. Our analysis provides just the first step of an important research problem that aims at enveilling the physical information encoded in correlation spreading and how this can be extracted in the next generation of experiments (see also \cite{kormos2017} for recent results in this direction).
In practice, the dynamics of the local maxima is easier to observe, and, as discussed above, our predictions are consistent with the existing observations.
Our analysis shows, however, that in generic experiments characterizing the spreading of correlations the data need be interpreted very carefully.
The propagation of local extrema
does not characterize the correlation edge.
Identifying the latter requires an accurate scaling analysis of the leaks.
Existing experimental data have been collected either in a regime of parameters where the two structures coincide~\cite{cheneau2012}, or on small systems where quantitative analysis is obfuscated by strong finite-size effects.
However the next generation of experiments based on Rydberg atoms, tampered wave-guides, and larger trapped-ion systems provide the natural setup to discern between the CE and the local features as our calculations suggest.

\acknowledgments
We thank Anton Buyskikh and Jad C.\ Halimeh for fruitful discussions.
This research was supported by the
European Commission FET-Proactive QUIC (H2020 grant No.~641122)
and the Deutsche Forschungsgemeinschaft (DFG) through SFB/CRC1073 (Projects~B03 and C03).

%%%%%%%%%%%%%%%%%%%%%%%%%%%%%%%%%%%%%%%%%%%%% 
     
\appendix

\section{Bose-Hubbard model}\label{sec:appBH}
The Bose-Hubbard (BH) model,
%%%%%%%%%%%%%%%%%%%%%%%%%%%%%%%%%%
\begin{equation}\label{eq:BHm}
\hat{H} = - J \sum_{\langle \R, \Rp \rangle} \left( \hat{a}_{\R}^\dagger \hat{a}_{\Rp} + \textrm{H.c.}\right) + \frac{U}{2}\sum_{\R} \hat{n}_{\R} \left( \hat{n}_{\R}-1 \right),
\end{equation}
%%%%%%%%%%%%%%%%%%%%%%%%%%%%%%%%%%
is constructed using the particle operators
$\hat{K}_1(\R) \equiv \hat{n}_{\R} \left(\hat{n}_{\R}-1\right)$ and
$\hat{K}_2(\R,\Rp) \equiv -\hat{a}_{\R}^\dagger\hat{a}_{\Rp}-\hat{a}_{\Rp}^\dagger\hat{a}_{\R}$,
where $\hat{a}_{\R}$ and $\hat{n}_{\R}=\hat{a}_{\R}^\dagger\hat{a}_{\R}$ are, respectively, the annihilation and number operators on the lattice site $\R$.
The amplitudes are, respectively, the two-body interaction strength, $h(\R)=U/2$, and the tunnel amplitude $J(\R,\Rp)=J$.
The Bose-Hubbard model has two phases, namely the superfluid (SF) phase for
$J \gg U$ and
the Mott-insulator (MI) phase for $J \ll U$. The precise critical point depends on the dimension and on the average number of particles per site. For a review, see for instance Ref.~\onlinecite{georges2010}.

\subsection{Superfluid phase}
In the superfluid phase and for high-enough average particle density in 1D, $n \gg U/2J$,
we may rely on Bogoliubov meanfield approximation. Assuming small density fluctuations, $\Delta n \ll n$, one develops the interaction term in Eq.~(\ref{eq:BHm}) up to quadratic order. The resulting quadratic form is then diagonalized using standard Bogoliubov transformation (see for instance Refs.~\onlinecite{natu2013,cevolani2015}). It yields the gapless dispersion relation
%%%%%%%%%%%%%%%%%%%%%%%%%%%%%%%%%%
\begin{equation}\label{eq:BHmSF.Ek}
E_k \simeq 2 \sqrt{2J\sin^{2}\left(k/2\right)\left[2J\sin^{2}\left(k/2\right)+nU\right]}.
\end{equation}
%%%%%%%%%%%%%%%%%%%%%%%%%%%%%%%%%%
It is phononic in the low-energy limit, $E_k \simeq c k$, with the sound velocity $c=\sqrt{2nJU}$. At higher energy, it shows an inflection point at some finite momentum $0<\kmax<\pi$, corresponding to the maximum group velocity $\Vgmax=\Vg(\kmax)$.

After the quench, the connected one-body correlation function, $G(R,t)=\langle \hat{a}^\dagger_R (t) \hat{a}_0 (t) \rangle - \langle \hat{a}^\dagger_R (0) \hat{a}_0 (0) \rangle$ considered in the main paper,
is then cast into the form of Eq.~(\ref{eq:gen-corr-tev}) by mapping the particle operators onto Bogoliubov quasi-particles operators. It yields the amplitude function
%%%%%%%%%%%%%%%%%%%%%%%%%%%%%%%%%%
\begin{equation}\label{eq:BHmSF.Fk}
\mathcal{F}(k) = \frac{2 n^2 J U_\textrm{f}(U_\textrm{f}-U_\textrm{i})}{E_{k,\textrm{i}} E_{k,\textrm{f}}^2} \sin^{2}({k}/{2}),
\end{equation}
%%%%%%%%%%%%%%%%%%%%%%%%%%%%%%%%%%
where the indices `i' and `f' refer to the pre-quench and post-quench values, respectively.

\subsection{Mott insulator phase}
In the Mott insulator phase, the model develops a finite gap. The energy excitations may be found using strong-coupling expansions, see for instance Refs.~\onlinecite{altman2002,huber2007,barmettler2012}.
For $n\in\mathbb{N}^*$ and $U \gg J n$, it yields low-energy excitations made of doublon-holon pairs of energy
%%%%%%%%%%%%%%%%%%%%%%%%%%%%%%%%%%
\begin{equation}\label{eq:BHmMI.Ek}
2E_{k} \simeq U - 2(2n+1)J\cos(k).
\end{equation}
%%%%%%%%%%%%%%%%%%%%%%%%%%%%%%%%%%
The maximum of the group velocity is at the center of the band,
$\kmax = \pi/2$, where the group and phase velocities are
$\Vgmax=(2n+1)J$ and $\Vphimax=U/\pi$, respectively. 
Hence, for $U>\pi(2n+1)J$, the phase velocity exceeds the group velocity,
$\Vphimax>\Vgmax$. Note that this regime is well inside the Mott regime where Eq.~(\ref{eq:BHmMI.Ek}) is accurate.

Similarly as for the superfluid phase, the connected one-body correlation function can be cast into the form of Eq.~(\ref{eq:gen-corr-tev}) with the amplitude function
%%%%%%%%%%%%%%%%%%%%%%%%%%%%%%%%%%
\begin{equation}\label{eq:BHmMI.Fk}
\mathcal{F}(k) = \frac{4 J n(n+1)}{i U} \sin(k),
\end{equation}
%%%%%%%%%%%%%%%%%%%%%%%%%%%%%%%%%%
for the quench from $U_\textrm{i} = \infty$ to $U_\textrm{f}=U$ as considered in the paper.

\section{Long-range XY and XXZ models}\label{sec:LRModels}
For spin models, the operators $\hat{K}_j$ represent spin operators,
the parameter $J(\R,\Rp)$ the exchange term, and $h(\R)$ a magnetic field.
For the long-range XY (LRXY) model, we use $\hat{K}_2(\R,\Rp) \equiv \hat{S}_{\R}^x\cdot\hat{S}_{\Rp}^x +\hat{S}_{\R}^y\cdot\hat{S}_{\Rp}^y$,
$J (\R,\Rp) = -J/2\vert\R-\Rp\vert^\alpha$,
and $h(\R)=0$.
For the initial state, it is generalized to the XXZ model by including an antiferromagnetic exchange coupling in the $z$ direction, which yields the Hamiltonian
%%%%%%%%%%%%%%%%%%%%%%%%%%%%%%%%%%
\begin{equation}\label{eq:XXZ.H}
 \hat{H} = \sum_{\R \neq \Rp} \frac{J/2}{\vert \R - \Rp \vert^\alpha} \left[ -\left( \hat{S}_{\R}^x \hat{S}_{\Rp}^x + \hat{S}_{\R}^y \hat{S}_{\Rp}^{y} \right) + \epsilon \hat{S}_{\R}^{z} \hat{S}_{\Rp}^{z}  \right].
\end{equation}
%%%%%%%%%%%%%%%%%%%%%%%%%%%%%%%%%%
For the LRXY case considered in the main paper, the quench is performed from the ground state of the XXZ model ($\epsilon \neq 0$) to the XY model ($\epsilon=0$).

We study the phase where the rotational symmetry around the $z$ axis is spontaneously broken and the spins are polarized along the $x$ axis.
There, the Hamiltonian can be diagonalized using standard Holstein-Primakoff transformation~\cite{holstein1940,auerbach1994},

%%%%%%%%%%%%%%%%%%%%%%%%%%%%%%%%%%
\begin{eqnarray}
S_{\R}^{x} & = &\frac{1}{2}-\hat{a}_{\R}^\dagger a_{\R},
\nonumber \\
S_{\R}^{y} & \simeq & -\frac{\hat{a}_{\R}^\dagger-\hat{a}_{\R}}{2 i},
\nonumber \\
S_{\R}^{z} & \simeq & -\frac{\hat{a}_{\R} + \hat{a}_{\R}^\dagger}{2},
\nonumber 
\end{eqnarray}
%%%%%%%%%%%%%%%%%%%%%%%%%%%%%%%%%%
where terms beyond second order in the boson operators $\hat{a}_{\R}$ and $\hat{a}_{\R}^\dagger$ are neglected.
Inserting these transformations into Eq.~(\ref{eq:XXZ.H}) yields a quadratic Bose Hamiltonian, which can be diagonalized using canonical Bogoliubov transformations,
see for instance Ref.~\onlinecite{frerot2017}.
For $\epsilon=0$ (LRXY model), it yields the dispersion relation for $D=1$
%%%%%%%%%%%%%%%%%%%%%%%%%%%%%%%%%%
\begin{equation}\label{eq:XXZ.Ek}
 E_{k} = \frac{J P_\alpha(0)}{2} \sqrt{1-\frac{P_\alpha (k)}{P_\alpha (0)}},
\end{equation}
%%%%%%%%%%%%%%%%%%%%%%%%%%%%%%%%%%
where $P_\alpha (k) = \int d R\ \e^{-i k\cdot R}/\vert R \vert^\alpha$ is the Fourier transform of the long-range term.
In the infrared limit, it can be written
%%%%%%%%%%%%%%%%%%%%%%%%%%%%%%%%%%
\begin{equation}\label{eq:Pinfrared}
P_\alpha(k) \approx P_\alpha(0) + P^\prime_\alpha k^{\alpha-D},
\end{equation}
%%%%%%%%%%%%%%%%%%%%%%%%%%%%%%%%%%
where $P_\alpha(0)$ and $P_\alpha^\prime$ are finite constants.
Hence, we find $E_k \propto \vert k \vert^z$ with $z=(\alpha-D)/2$.
For $D < \alpha < D+2$, the quasi-particle energy is finite but the group velocity $\Vg$ diverges in the infrared limit $k \rightarrow 0$.

The connected spin-spin correlation function along the $z$ direction for a quench from $\epsilon_\textrm{i}\neq 0$ to $\epsilon_\textrm{f}=0$,
$G_0(R,t) = \langle S_R^z(t) S_0^z(t) \rangle - \langle S_R^z(t)\rangle \langle S_0^z(t) \rangle$,
used in the paper is cast into the form of Eq.~(\ref{eq:gen-corr-tev}) using the quasi-particle amplitudes, which yields
%%%%%%%%%%%%%%%%%%%%%%%%%%%%%%%%%%
\begin{equation}\label{eq:LRXY.Fk}
 \mathcal{F}\left(k\right)=\frac{\epsilon_\textrm{i}}{8}\frac{P_\alpha\left(k\right)}{P_\alpha\left(0\right)}\sqrt{\frac{P_\alpha(0)-P_\alpha\left(k \right)}{P_\alpha(0)+\epsilon_\textrm{i}P_\alpha\left(k\right)}}.
\end{equation}
%%%%%%%%%%%%%%%%%%%%%%%%%%%%%%%%%%
In the infrared limit, it scales as $\mathcal{F}\left( k \right) \sim k^\nu$ with $\nu=(\alpha-D)/2=z$.

The linearization of the Holstein-Primakoff transformation holds for $\vert 1/2 - S^x_\R \vert \ll 1$, see Ref.~\onlinecite{auerbach1994}. For the calculations corresponding to Fig.~\ref{fig:spins}~(a), we find $\max\{\vert 1/2 - S^x_\R \vert \} \simeq 0.12$. It validates the spin-wave approximation used in the paper. This result agrees with the predictions for the same model made in Ref.~\onlinecite{frerot2017} where the validity of the spin wave approach for that model is extensively studied.

\section{Long-range transverse Ising model}
The long-range transverse Ising (LRTI) model corresponds to the spin operators
$\hat{K}_1(\R) \equiv \hat{S}_{\R}^z$
and $\hat{K}_2(\R,\Rp) \equiv \hat{S}_{\R}^x\cdot\hat{S}_{\Rp}^x$
with a uniform magnetic field $h(\R)=-2h$
and the algebraically decaying exchange amplitude $J (\R,\Rp) = 2J/\vert\R-\Rp\vert^\alpha$,
which yields
%%%%%%%%%%%%%%%%%%%%%%%%%%%%%%%%%%
\begin{equation}\label{eq:LRTI}
\hat{H} =  \sum_{\R \neq \Rp} \frac{2J}{\vert \R - \Rp \vert^\alpha} \hat{S}_{\R}^{x}\hat{S}_{\Rp}^{x} - 2h \sum_{\R} \hat{S}_{\R}^{z}.
\end{equation}
%%%%%%%%%%%%%%%%%%%%%%%%%%%%%%%%%%
The LRTI has two phases~\cite{koffel2012}. In the $z$-polarized phase, the disprsion relation can be found using again the Holstein-Primakoff transformation
%%%%%%%%%%%%%%%%%%%%%%%%%%%%%%%%%%
\begin{eqnarray}
S_{\R}^{x}  & \simeq & \frac{\hat{a}_{\R} + \hat{a}_{\R}^\dagger}{2},
\nonumber \\
S_{\R}^{y} & \simeq & -\frac{\hat{a}_{\R}^\dagger-\hat{a}_{\R}}{2 i},
\nonumber \\
S_{\R}^{z} & = &\frac{1}{2}-\hat{a}_{\R}^\dagger a_{\R},
\nonumber 
\end{eqnarray}
%%%%%%%%%%%%%%%%%%%%%%%%%%%%%%%%%%
One then finds the dispersion relation for $D=1$
%%%%%%%%%%%%%%%%%%%%%%%%%%%%%%%%%%
\begin{equation}
E_k = 2 \sqrt{h \left[ h + J P_\alpha (k) \right]}.
\end{equation}
%%%%%%%%%%%%%%%%%%%%%%%%%%%%%%%%%%
In the infrared limit, where Eq.~(\ref{eq:Pinfrared}) holds, it can be expressed as
%%%%%%%%%%%%%%%%%%%%%%%%%%%%%%%%%%
\begin{equation}
E_k = \Delta + c \vert k\vert^z,
\end{equation}
%%%%%%%%%%%%%%%%%%%%%%%%%%%%%%%%%%
where the gap
$\Delta = 2 \sqrt{h \left[ h + J P_\alpha (0) \right]}$ is finite,
$c = \sqrt{\frac{h}{h+JP_\alpha(0)}}JP_\alpha^\prime$,
and
$z=\alpha-D$,
see Ref.~\onlinecite{cevolani2016}.
Hence, the quasi-particle energy is finite and the group velocity diverges for $D<\alpha<D+1$.

In the main paper, we consider the connected spin-spin correlation function along the $x$ direction,
$G_0(R,t) = \langle S_R^x(t) S_0^x(t) \rangle - \langle S_R^x(t)\rangle \langle S_0^x(t) \rangle$, which can be written in the form of Eq.~(\ref{eq:gen-corr-tev}) with
%%%%%%%%%%%%%%%%%%%%%%%%%%%%%%%%%%
\begin{equation}\label{eq:LRTI.Fk}
 \mathcal{F}(k)=\frac{h\left(J_\textrm{i}-J_\textrm{f}\right)P_\alpha \left(k\right)}{8\left[h+J_\textrm{f}P_\alpha\left( k \right)\right]\sqrt{h \left[ h + J_\textrm{i} P_\alpha (k) \right]}}.
\end{equation}
%%%%%%%%%%%%%%%%%%%%%%%%%%%%%%%%%%
In the infrared limit, it converges to a finite value. Hence $\mathcal{F}\left(k\right) \sim k^\nu$ with $\nu=0$.

For the calculations corresponding to the Ising model in Fig.~\ref{fig:spins}~(b), we find $\max\{\vert 1/2 - S^z_\R \vert\}=0.11$, which validates the spin-wave approximation also in this case.

\section{Numerical analysis of the local extrema and the correlation edge for the LRXY and LRTI models}\label{sec:Edgeextraction}
In the main paper, it is shown that the causality cone features a double structure: an outer structure, which determines the correlation edge (CE), and an
inner structure where local extrema propagate. Here we provide details on the numerical analysis of the CE and of the trajectory for the first local extrema, for both the LRXY and LRTI models considered in the main manuscript.

\subsection{LRXY model}

\begin{figure}[t!]
\includegraphics[width=\columnwidth ]{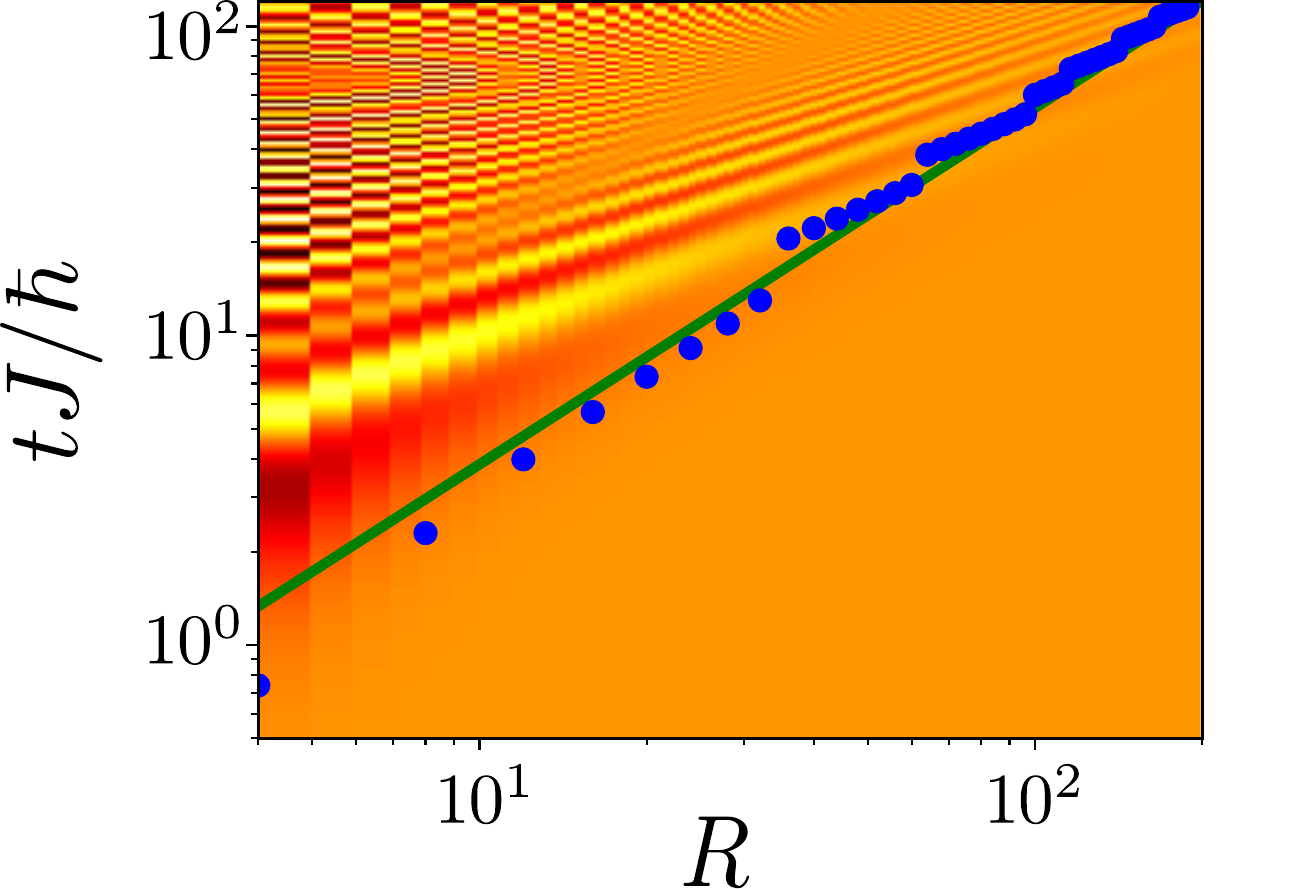}
\footnotesize{\caption{\label{fig:lr_edge_XY}
Spreading of the connected spin-spin correlation function for the LRXY model with $\alpha=2.3$ (same data as in Fig.~\ref{fig:spins}~(a) of the main paper; log-log scale).
The filled blue points correspond for each distance $R$ to the first time where the correlation reaches $2\%$ of its maximum value.
The solid green line shows the power law predicted theoretically with a fitted multiplicative factor.  
}}
\end{figure}

For the LRXY model, we consider the time evolution of the connected spin-spin correlation function $G \left( R, t \right)$ along the $z$ axis. 
Figure~\ref{fig:lr_edge_XY} shows the same data as Fig.~\ref{fig:spins}~(a) in the main paper.
To find the CE, we proceed as follows. For each distance $R$, we trace the activation time $t^\star(R)$ corresponding to the first
time when a fraction ($2\%$) of the absolute maximum of the correlation function is reached.
It yields the filled blue points on Fig.~\ref{fig:lr_edge_XY}. The latter feature a linear trajectory in the log-log scale of the figure, that is a power law behavior in lin-lin scale. The latter is in excellent agreement with the theoretical prediction $\betaLR \simeq 1.15$ shown as a solid green line on the figure.
We have also fitted a power law function, $t^\star \propto R^{\betaLRfit}$, to the blue points for $20 < R < 175$ (not shown on the figure). It yields
$\betaLRfit = 1.083 \pm 0.013$ in good agreement with the prediction.

\begin{figure}[t!]
\includegraphics[width=\columnwidth ]{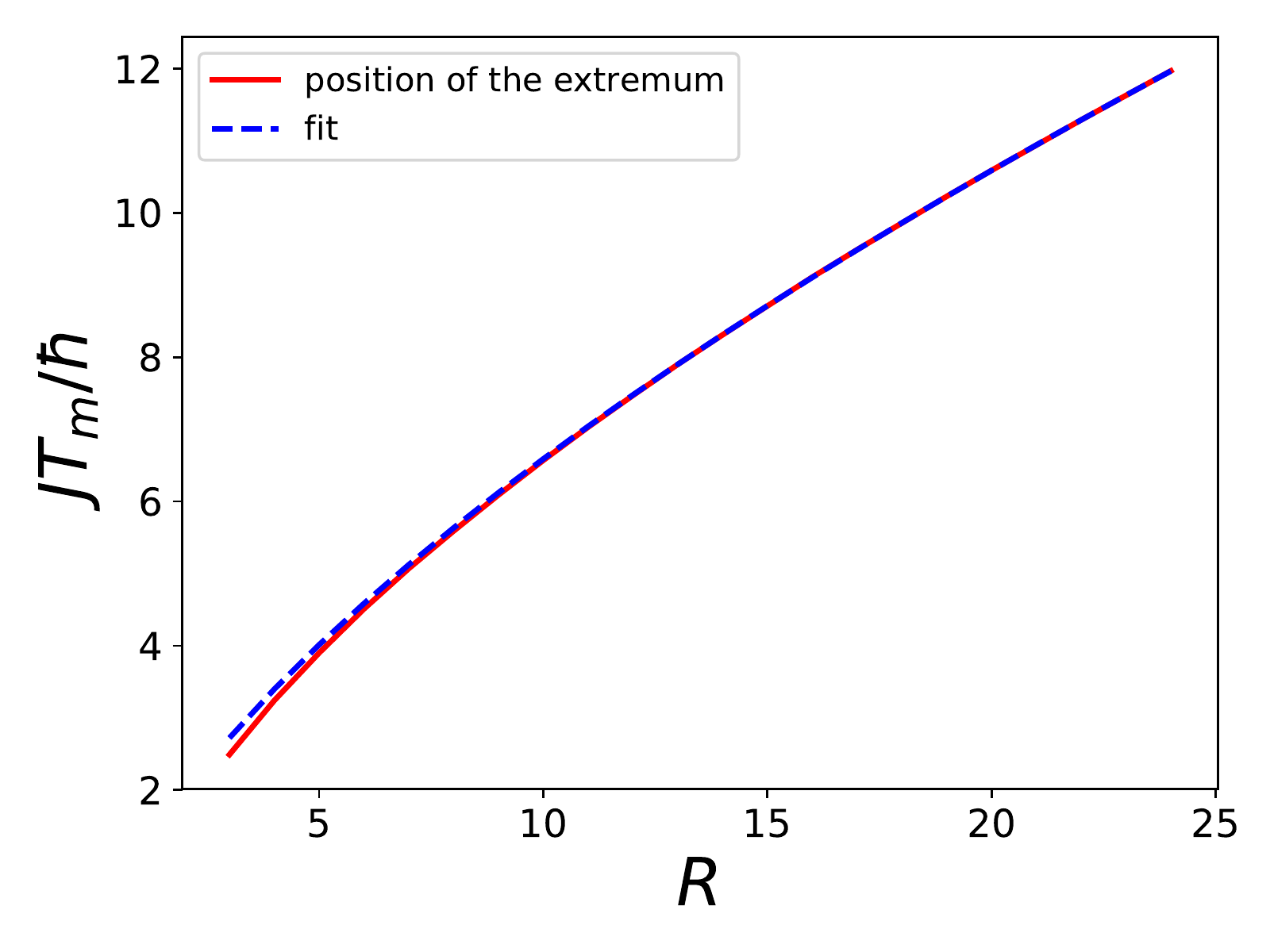}
\footnotesize{\caption{\label{fig:maxima_XY}
Trajectory of the first extremum of the connected spin-spin correlation function for the LRXY model (lin-lin scale).
The figure shows the numerical results found from Fig.~\ref{fig:lr_edge_XY} (solid red line) together with a fitted power law (dashed blue line).}}
\end{figure}

A similar result is obtained on a length scale closer to the one accessible in state-of-the-art experiments~\cite{zhang2017}. Fitting the same algebraic function to the correlation edge in the range $10<R<30$ and $t<10/J$ yields $\betaLRfit=1.121\pm 0.012$. It is already in good agreement with the fit in the larger range and with the theoretical prediction.

To analyze the behaviour of the local extrema, we trace them from the data of Fig.~\ref{fig:lr_edge_XY} where they are clearly visible. The result for the first one is plotted on Fig.~\ref{fig:maxima_XY} (solid red line) together with a fitted power law, $t_{\mathrm{m}} = a R^{\betamfit} + b$ (dashed blue line). The fit yields $\betamfit = 0.634 \pm 0.014$ in excellent agreement with the theoretical exponent $\betam = 0.65$.

\subsection{LRTI model}

\begin{figure}[t!]
\includegraphics[width=\columnwidth]{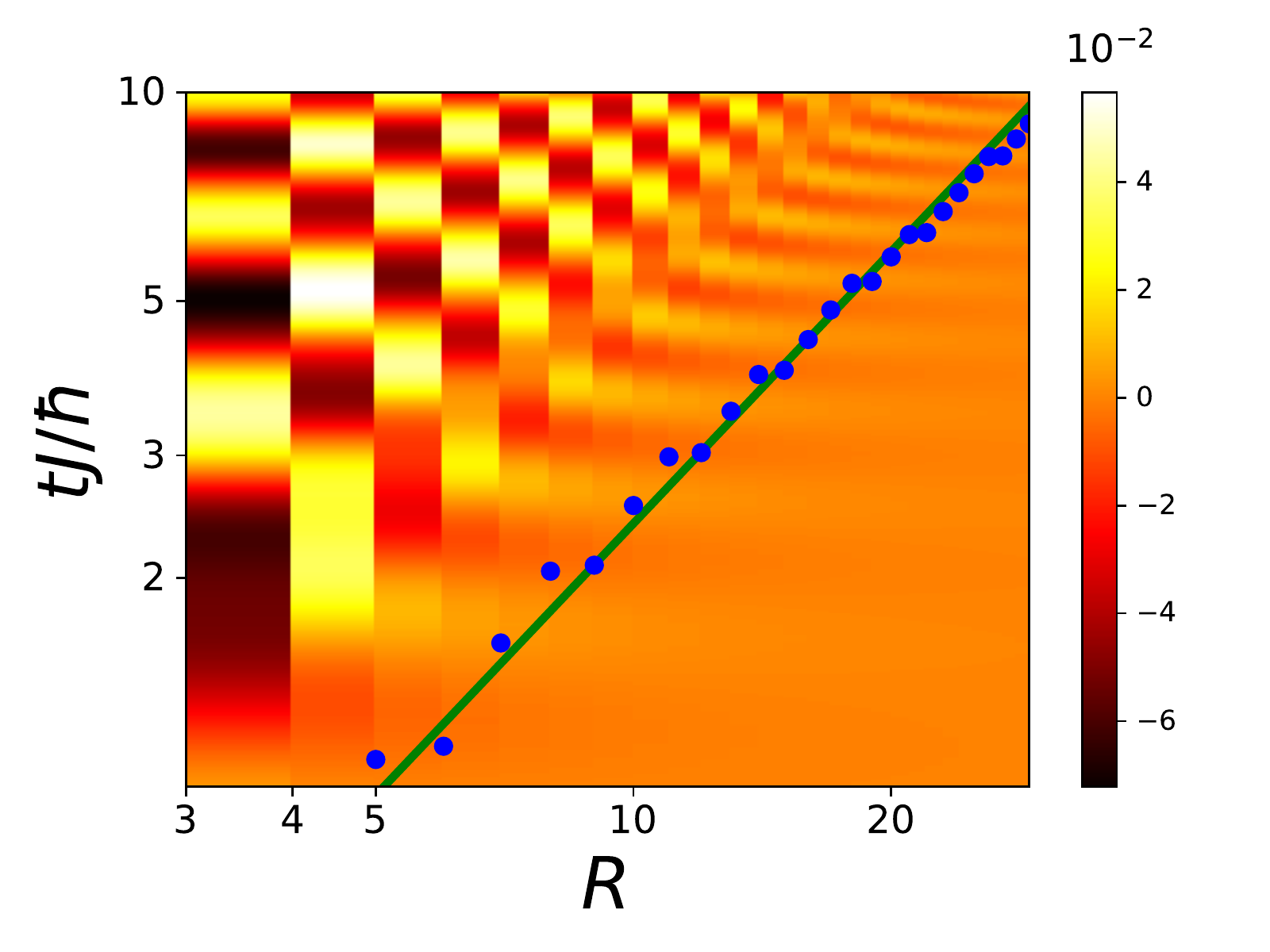}
\footnotesize{\caption{\label{fig:lr_edge_TI}
Spreading of the connected spin-spin correlation function for the LRTI model model with $\alpha=1.7$ (same data as in Fig.~\ref{fig:spins}(b) of the main paper; log-log scale).
The filled blue points correspond for each distance $R$ to the first time where the correlation reaches $2.8\%$ of its maximum value.
The solid green line shows the power law predicted theoretically with a fitted multiplicative factor.
}}
\end{figure}

\begin{figure}[t!]
\includegraphics[width=\columnwidth]{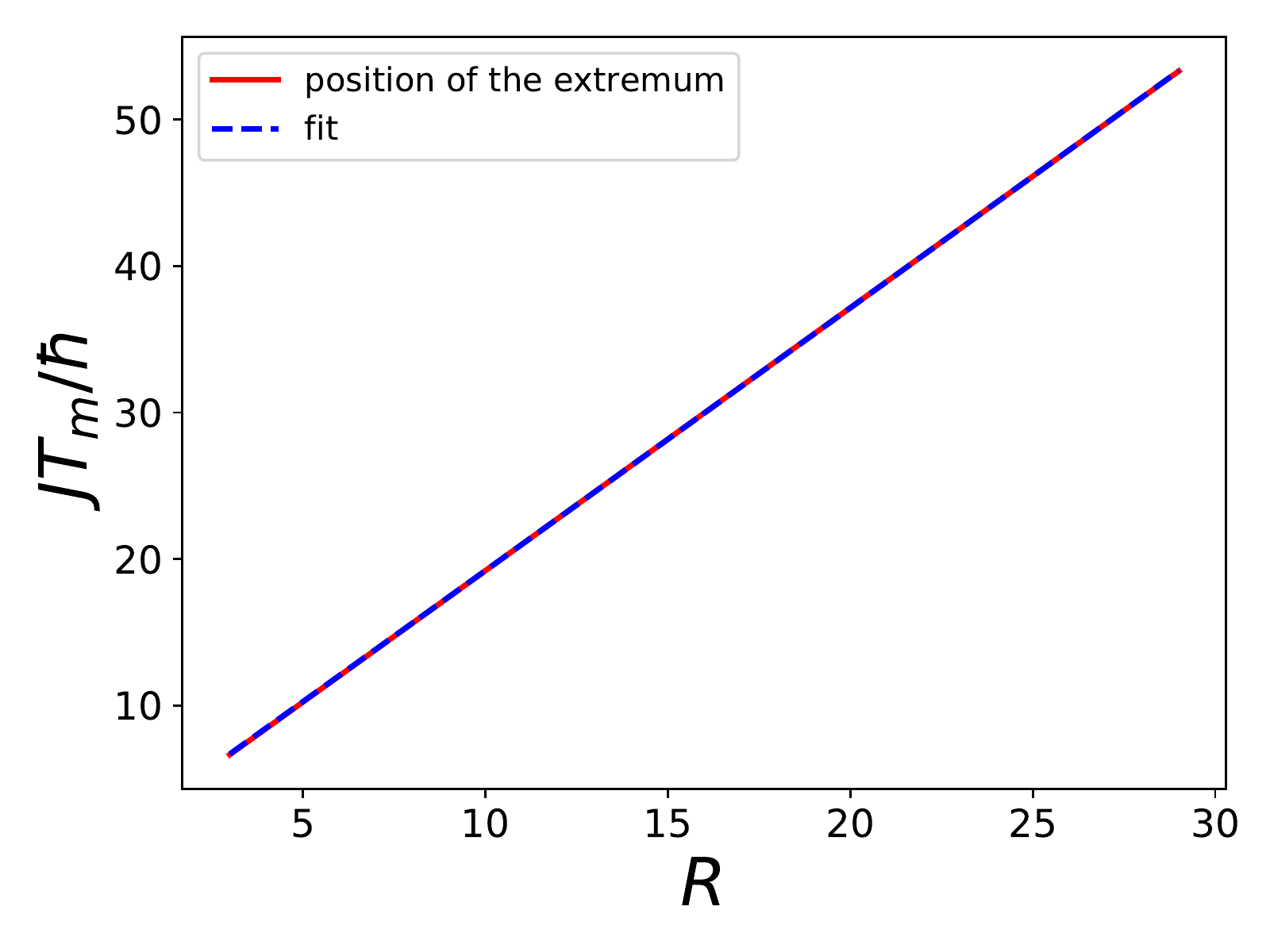} 
\footnotesize{
\caption{\label{fig:maxima_TI}
Trajectory of an extremum of the connected spin-spin correlation function for the LRTI model (lin-lin scale).
It corresponds to the dashed, dark blue line in Fig.~\ref{fig:spins}(b) of the main paper.
The figure shows the numerical results found from Fig.~\ref{fig:lr_edge_TI} (solid red line) together with a fitted power law (dashed blue line).
}}
\end{figure}

We now turn to the LRTI model and perform the same analysis as above, up to details that we discuss below.
Figure~\ref{fig:lr_edge_TI} shows the same data as Fig.~\ref{fig:spins}(b) in the main paper but on a smaller time scale.
The CE, corresponding to the trajectory of the first points where the correlation function reaches a fraction
($2.8\%$) of its maximum (filled blue points) matches very well the theoretical prediction
$t^{*} \propto R^{\betaLR}$ with $\betaLR = 1.3$ (solid green line with a fitted multiplicative factor).
Moreover, fitting a power law function to the blue points for $4 < R < 30$ yields $\betaLRfit = 1.28 \pm 0.02$ (not shown on the figure),
in perfect agreement with our prediction.

We analyse the trajectory of the maxima in the same way as before. For the LRTI model, ballistic spreading is expected sufficently large values of $t$ and $R$.
This is confirmed by the behaviour of the local maxima in Fig.~\ref{fig:spins}(b) of the main paper. The first local maxima are marked by dashed light blue curves.
In the figure, they are curved owing to the competition between the different terms of the phase of the cosine in Eq.~(\ref{eq:statphaseLR}) of the main manuscript.
They, however, clearly converge to a ballistic behaviour when $t$ and $R$ increase.
For the range of $t$ and $R$ presented in Fig.~2 (b) the dark blue line is purely ballistic even fro small values of $t$ and $R$. 
Its trajectory is shown in lin-lin scale on Fig.~\ref{fig:maxima_TI}.
To confirm the ballistic behaviour, we fitted the power-law function $t_{\textrm{m}}=aR^{\betamfit}+b$ to the date (dashed blue line on the figure).
The fit gives $\betamfit=1.0045\pm0.0003$ in excellent agreement with the theoretical exponent $\betam=1$.

We have also performed similar fits on the other local maxima, which confirms the asymptotic ballistric behaviour. For instance, for the first local maximum (lowest dashed, light-blue lines in Fig.~\ref{fig:spins}(b) of the main paper) and $10<R<20$, we find $\betamfit=1.0039\pm 0.0002$, which is also in excellent agreement with the theoretical prediction.

%%%%%%%%%%%%%%%%%%%%%%%%%%%%%%%%%%%%%%%%%%%%

\end{document}